\documentclass[conference]{IEEEtran}
\IEEEoverridecommandlockouts

\usepackage{cite}
\usepackage{amsmath,amssymb,amsfonts}
\usepackage{graphicx}
\usepackage{textcomp}
\usepackage{algorithm}
\usepackage{algpseudocode}
\usepackage{array}
\usepackage{booktabs}
\usepackage{url}
\usepackage{tabularx}
\usepackage{adjustbox}
\usepackage{hhline}
\usepackage[table]{xcolor}
\usepackage{booktabs}    
\usepackage{pifont}      
\usepackage{graphicx}    
\usepackage{multirow}

\def\BibTeX{{\rm B\kern-.05em{\sc i\kern-.025em b}\kern-.08em
    T\kern-.1667em\lower.7ex\hbox{E}\kern-.125emX}}

\title{Mobicom-26 UPF}
\author{ 
\IEEEauthorblockN{Padmapriya Patil$^1$, Abhishek Bhattacharyya$^1$, Gudepu Venkateswarlu$^1$, Andrea Fumagalli$^1$, Koteswararao Kondepu$^2$
}
\IEEEauthorblockA{
$^1$ Open Networking Advanced Research (OpNeAR) Lab, The University of Texas at Dallas, TX, USA\\
}
\IEEEauthorblockA{
$^2$ Computer Science and Engineering, Indian Institute of Technology Dharwad, Dharwad, 580011, KA, India
}
\IEEEauthorblockA{E-mail: Padmapriya.Patil@UTDallas.edu}
}

\title{Implementation of Hybrid QoS-Aware Data Radio Bearer in 5G Networks}
\date{February 2026}

\begin{document}

\maketitle

\begin{abstract}

In Fifth-Generation~(5G) wireless networks, Quality of Service~(QoS) flow management supports diverse applications with heterogeneous QoS requirements, including latency-sensitive and high-throughput services. 5G architectures adopt a flow-based QoS framework, where QoS flows are mapped to Data Radio Bearers~(DRBs) to achieve the desired radio resource allocation. Existing mapping strategies typically rely on static one-to-one QoS-to-DRB mappings per PDU session, leading to inefficient spectrum utilization or inadequate support for delay-sensitive traffic.

This work presents a hybrid QoS-aware DRB mapping mechanism based on standardized 5G QoS Identifier~(5QI) characteristics. The proposed framework incorporates multiple QoS flows within a single PDU session, supporting both many-to-one and one-to-one mapping strategies. The hybrid framework improves radio resource utilization and reduced bearer overhead by enabling flexible DRB utilization, where compatible non-critical flows share DRBs while delay-critical flows are assigned dedicated DRBs to ensure strict QoS guarantees. The proposed hybrid QoS-aware DRB mapping is implemented using the open-source OpenAirInterface~(OAI) mobile stack platform to demonstrate the integration of the proposed features within the existing 5G framework.
    
\end{abstract}


\section{Introduction}

The Fifth-generation (5G) network supports a wide range of emerging applications, including ultra-high-definition video streaming, augmented and virtual reality, industrial automation, and mission-critical communications, each with distinct Quality of Service~(QoS) requirements. To accommodate these diverse demands, the 5G New Radio~(NR) architecture introduces a flow-based QoS framework, where each QoS flow is identified by a QoS Flow Identifier~(QFI) and characterized using standardized 5QI parameters~\cite{3gpp_ts23501_v16_6_0}. The Service Data Adaptation Protocol~(SDAP) maps QoS flows to Data Radio Bearers~(DRBs), while the User Plane Function~(UPF) classifies traffic and the gNB enforces QoS through radio resource allocation and scheduling.
Several studies have investigated QoS classification and mapping. Statistical traffic classification for Class of Service (CoS) mapping was proposed in~\cite{roughan2004class}, improving application identification without addressing radio resource allocation. QoS-MAN~\cite{satka2022qos} maps Time-Sensitive Networking (TSN) traffic to 5G QoS flows based on latency, jitter, and bandwidth requirements; however, it focuses on inter-network QoS translation rather than radio resource management. Similarly, the semantic-aware QoS framework in~\cite{zhang2021towards} enables cross-layer interaction but does not address QoS Flow-to-radio resource mapping.
Despite these advances, current implementations typically employ a single QoS flow-to-DRB mapping strategy per PDU session~\cite{oai_openairinterface5g}, limiting flexibility under dynamic and heterogeneous traffic conditions. To address this limitation, this paper proposes a hybrid QoS flow-to-DRB mapping mechanism that combines many-to-one and one-to-one mapping based on standardized 5QI characteristics. GBR and Non-GBR QoS flows with compatible QoS requirements are mapped to a shared DRB to improve radio resource utilization and reduce bearer overhead, while Delay-Critical GBR and custom QoS flows are assigned dedicated DRBs to preserve strict QoS guarantees. The proposed framework also supports adaptive isolation for delay-critical services when required.
The main contributions of this work are:
\begin{itemize}
\item Design of a hybrid QoS-aware DRB mapping mechanism based on 5QI classification.
\item Development of an intelligent 5QI-based QoS flow grouping strategy.
\item Experimental implementation and evaluation using OpenAirInterface (OAI).
\end{itemize}

\begin{figure}[!htb]
 \vspace{-0.5cm}
    \centering
\includegraphics[width=0.9\columnwidth]{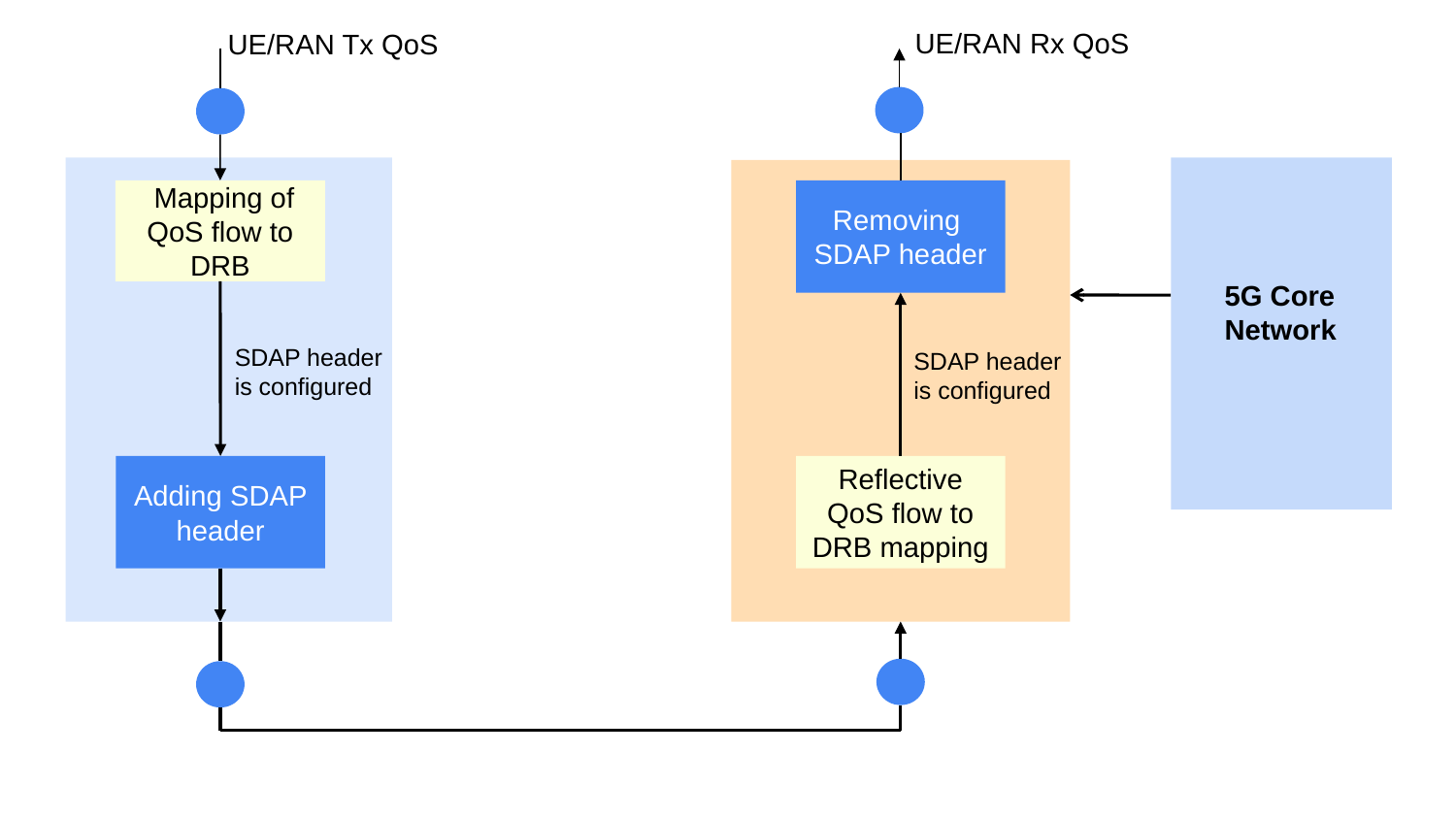}
    \vspace{-0.8cm}
    \caption{QoS to DRB mapping}
    \label{fig:qos_mapping}
     \vspace{-0.3cm}
\end{figure}

\section{Background}
This section provides a brief description of managing QoS Flows, DRB flow, and dynamic QoS flow in 5G networks.

\subsection{Managing QoS Flows}
QoS is a key mechanism in modern communication networks that provides differentiated treatment of traffic according to application requirements. Services such as ultra-high-definition video streaming, industrial automation, and mission-critical communications impose diverse latency, reliability, throughput, and packet loss constraints. Unlike LTE, which manages QoS at the bearer level using fixed QoS Class Identifiers (QCIs), 5G adopts a flow-based QoS architecture that enables fine-grained service differentiation.
Within a PDU session, multiple QoS flows can coexist, each identified by a QFI and characterized by standardized 5QI parameters, including priority level, packet delay budget, and packet error rate~\cite{3gpp_ts23501_v16_6_0}. During PDU session establishment, the SMF creates QoS flows and assigns QFIs, which are communicated to the UE and gNB. The UPF classifies packets according to policy rules and inserts the corresponding QFI into the GTP-U extension header. At the gNB, the SDAP layer maps QFIs to DRBs, while the PDCP, RLC, and MAC layers enforce reliability and scheduling according to the associated QoS requirements. This flow-based framework enables end-to-end QoS enforcement, network slicing, and application-aware traffic management.
\subsection{DRB Mapping}
QoS-to-DRB mapping translates logical QoS requirements into radio-layer resources. During PDU Session Resource Setup, the gNB associates one or more QoS flows with DRBs through SDAP configuration. Multiple QoS flows may share a DRB (many-to-one mapping) to improve radio resource utilization, whereas latency-sensitive or isolated services may use dedicated DRBs (one-to-one mapping). As illustrated in Fig.~\ref{fig:qos_mapping}, SDAP performs the QFI-to-DRB association, while the lower protocol layers implement scheduling and retransmission policies according to the QoS characteristics.
\subsection{QoS Flow in the Core Network}
Within the Core Network, the PCF defines QoS policies using standardized 5QI parameters and delivers them to the SMF through Policy and Charging Control (PCC) rules. The SMF configures the UPF using PFCP signaling, where Packet Detection Rules (PDRs), Forwarding Action Rules (FARs), and QoS Enforcement Rules (QERs) determine packet classification, forwarding behavior, and QFI assignment. The UPF inserts the appropriate QFI into the GTP-U extension header for downlink packets, enabling the gNB to apply the corresponding radio-layer QoS treatment. This policy-driven framework provides consistent end-to-end QoS enforcement across the core and radio access network.
\subsection{Dynamic QoS Allocation in 5G}
5G supports runtime modification, addition, and removal of QoS flows within an active PDU session without re-establishing the session. When application requirements or network policies change, the SMF updates the QoS configuration and initiates a PFCP Session Modification toward the UPF. Corresponding QoS updates are conveyed to the gNB through NGAP signaling, allowing the SDAP layer to modify QoS flow-to-DRB associations or create new DRBs when required. This dynamic framework enables application-aware QoS adaptation while maintaining session continuity and efficient radio resource utilization.

\begin{figure*}[!htb]
\centering

\begin{minipage}{0.524\textwidth}
    \centering    \includegraphics[width=\linewidth]{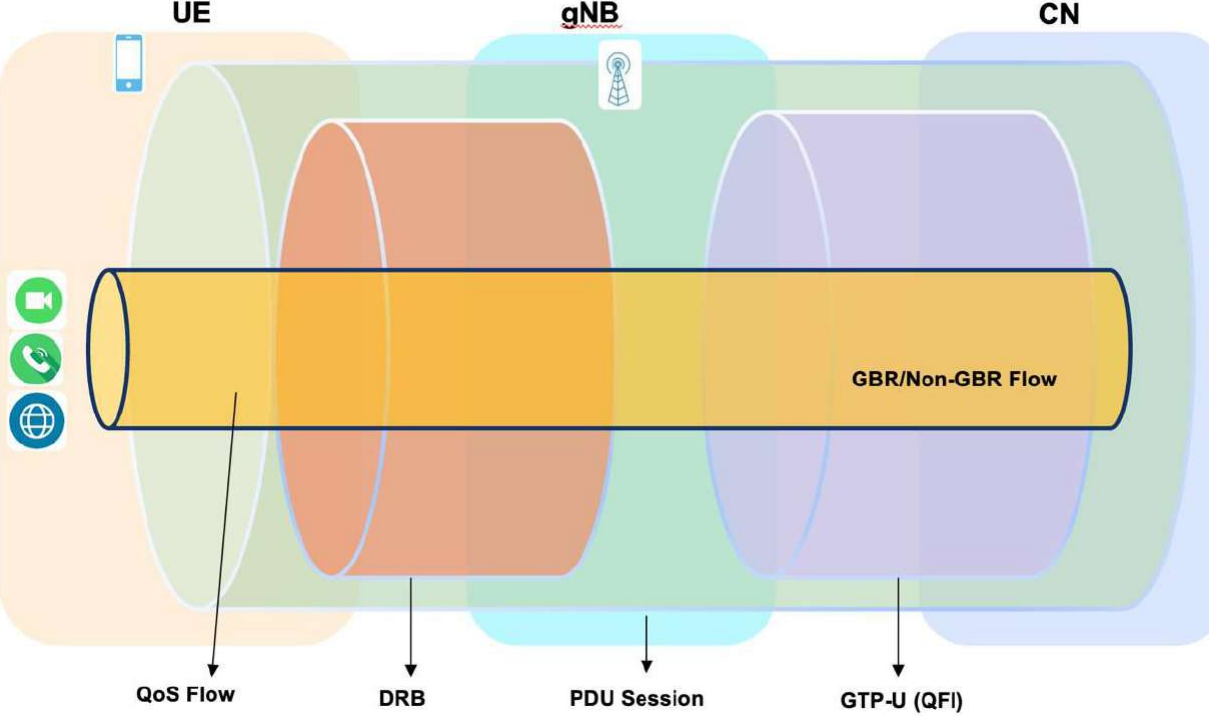}
    \caption{Baseline Architecture}
    \label{fig:baseline}
\end{minipage}
\hfill
\begin{minipage}{0.47\textwidth}
    \centering
\includegraphics[width=\linewidth]{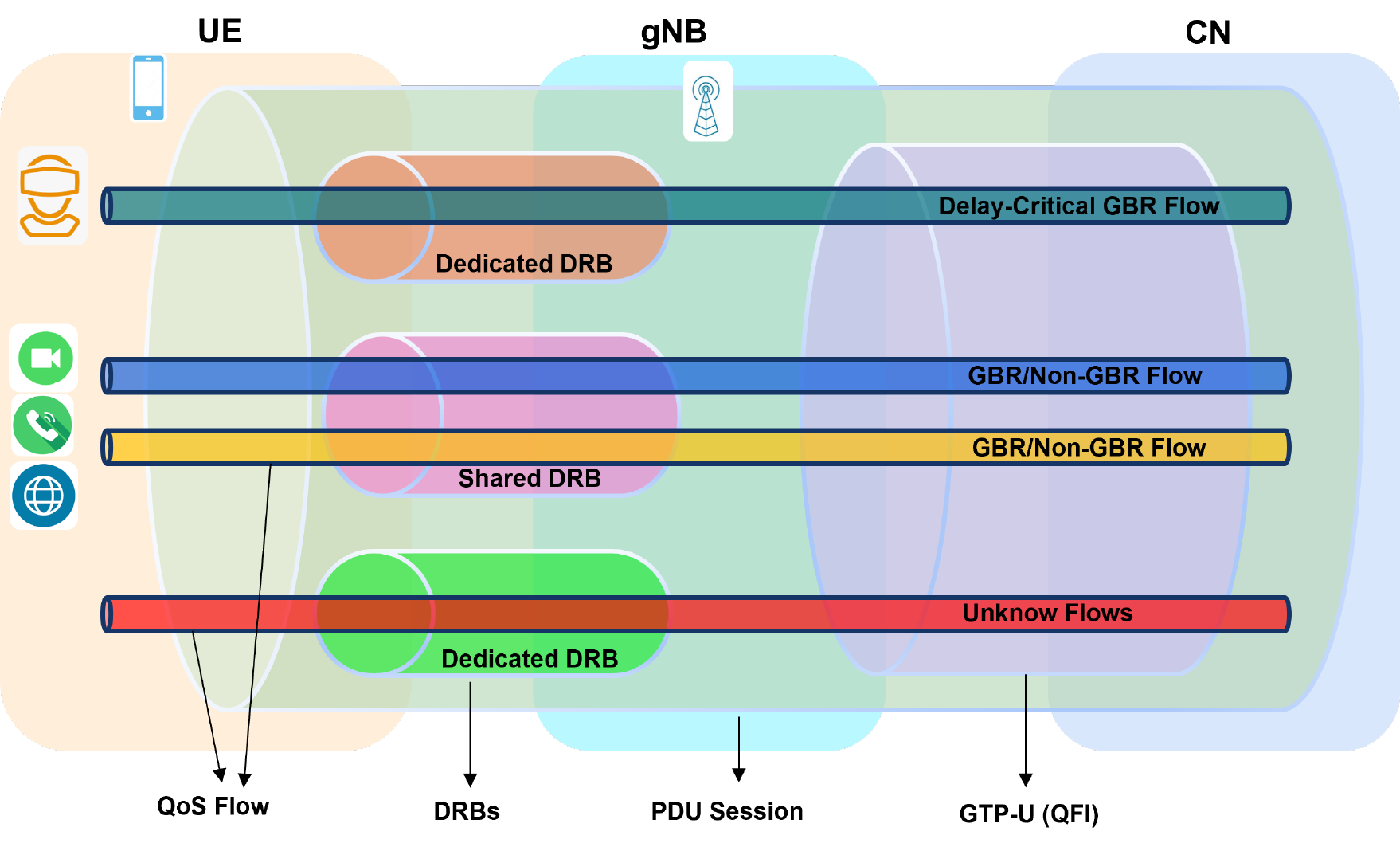}
    \caption{System Architecture: Proposed Hybrid Mapper}
    \label{fig:sys-arch}
\end{minipage}

\end{figure*}

\section{System Architecture and Proposed Approach}
\label{sec:sys-arch-prop-approach}

\subsection{Proposed Approach}
\label{sec:proposed-approach}
\begin{table}[!t]
\centering
\caption{Description of Variables in Algorithm~\ref{alg:qfi_drb_mapping2}}
\label{tab:variables}
\begin{tabular}{p{0.25\columnwidth} >{\raggedright\arraybackslash}p{0.68\columnwidth}}
\toprule
\textbf{Symbol} & \textbf{Description} \\
\midrule

$\mathcal{Q}$ & Set of QoS flows $[q_1,\dots,q_n]$ \\

$q_i$ & Individual QoS flow \\

$dc$ & Delay-critical mode flag \\

$\mathcal{D}$ & Set of DRBs \\

$d_i$ & Dedicated DRB assigned to $q_i$ \\

$d_s$ & Shared DRB \\

$\mathcal{S}$ & Set of shared QoS flows \\

$N_{\text{shared}}$ & Number of shareable flows \\

$N_{\text{dc}}$ & Number of delay-critical flows \\

$N_{\text{unknown}}$ & Number of unclassified flows \\

$N_{\text{DRB}}$ & Total number of required DRBs \\

5QI & QoS Classification Identifier \\

\bottomrule
\end{tabular}
\vspace{-0.3cm}
\end{table}

The current OAI implementation associates all QoS Flows within a PDU session with a single DRB, irrespective of their QoS characteristics. As illustrated in Fig.~\ref{fig:baseline}, GBR, Non-GBR, and Delay-Critical traffic share the same DRB, which simplifies bearer management but may increase contention among heterogeneous flows. This limitation motivates the proposed QoS-aware mapping mechanism.

The proposed hybrid framework, shown in Fig.~\ref{fig:sys-arch}, dynamically maps QoS Flows to DRBs according to their standardized 5QI characteristics. Algorithm~\ref{alg:qfi_drb_mapping2} takes as input the set of QoS flows $\mathcal{Q}=\{q_1,q_2,\dots,q_n\}$ and the Delay-Critical mode flag $dc$, and returns the final QoS Flow-to-DRB mapping. Depending on the value of $dc$ and the QoS requirements, the framework applies either one-to-one dedicated mapping or many-to-one shared mapping.

An empty DRB set $\mathcal{D}$ is initialized to store the DRBs created during algorithm execution (line~1). The algorithm then checks whether DC mode is enabled (line~2). This condition has the highest priority because DC mode enforces strict isolation among QoS flows.

When DC mode is enabled (lines~3--7), a new DRB $d_i$ is created for each QoS flow $q_i\in\mathcal{Q}$. The flow is mapped to the corresponding DRB, which is then inserted into $\mathcal{D}$. This produces a one-to-one QoS Flow-to-DRB mapping. After all flows have been processed, $\mathcal{D}$ is returned (line~8), and the algorithm terminates without further classification (line~9).

When DC mode is disabled, an empty shared set $\mathcal{S}$ is initialized (line~10). Three counters, $N_{\text{shared}}$, $N_{\text{dc}}$, and $N_{\text{unknown}}$, are also initialized (line~11) to record the numbers of shareable, Delay-Critical, and unknown QoS flows, respectively.

The algorithm iterates over all QoS flows in $\mathcal{Q}$ (line~12). Each flow is classified according to its standardized 5QI value (line~13), following the QoS characteristics defined in~\cite{3gpp_ts23501_v16_6_0}.

GBR and Non-GBR flows are added to the shared set $\mathcal{S}$, and $N_{\text{shared}}$ is incremented (lines~14--16). These flows can be carried by one shared DRB, reducing the number of independently configured bearers and improving radio-resource utilization.

Delay-Critical GBR flows are not added to $\mathcal{S}$. Instead, $N_{\text{dc}}$ is incremented (lines~17--18), since such flows require dedicated DRBs to preserve traffic isolation and stringent latency requirements.

Flows that do not match any predefined 5QI class are categorized as unknown, and $N_{\text{unknown}}$ is incremented (lines~19--20). These flows are conservatively assigned dedicated DRBs because their QoS requirements cannot be safely inferred.

The total number of required DRBs is then determined (line~23). If at least one shareable flow exists, the total consists of one shared DRB together with the dedicated DRBs required for Delay-Critical and unknown flows. Otherwise, only dedicated DRBs are required.

The shared DRB is created in lines~24--29. If $N_{\text{shared}}>0$, a shared DRB $d_s$ is created, and every QoS flow in $\mathcal{S}$ is mapped to it. The shared DRB is then inserted into $\mathcal{D}$, thereby implementing the many-to-one mapping policy.

Dedicated DRBs are subsequently created for Delay-Critical and unknown QoS flows (lines~31--36). For each such flow $q_i$, a new DRB $d_i$ is created, the flow is mapped to it, and the DRB is added to $\mathcal{D}$. This step preserves isolation for flows with strict or unclassified QoS requirements.

Finally, the completed DRB set $\mathcal{D}$ is returned (line~38). Thus, every QoS flow is assigned either to the shared DRB or to a dedicated DRB.

Overall, the algorithm applies one-to-one mapping when DC mode is enabled. Otherwise, it groups GBR and Non-GBR flows into a shared DRB while assigning Delay-Critical GBR and unknown flows to dedicated DRBs. This hybrid approach provides bearer sharing and traffic isolation within the same PDU session.

\begin{algorithm}[!htb]
\caption{Hybrid QoS Flow to DRB Mapping}
\label{alg:qfi_drb_mapping2}
\begin{algorithmic}[1]

\Require Set of QoS flows $\mathcal{Q}=\{q_1,q_2,\dots,q_n\}$, DC mode flag $dc$
\Ensure Mapping of QoS flows to DRBs

\State Initialize empty DRB list $\mathcal{D} \gets \emptyset$

\If{$dc = \text{true}$}
    \For{each $q_i \in \mathcal{Q}$}
        \State Create DRB $d_i$
        \State Map $q_i \rightarrow d_i$
        \State $\mathcal{D} \gets \mathcal{D} \cup \{d_i\}$
    \EndFor
    \State \Return $\mathcal{D}$
\EndIf

\State Initialize shared set $\mathcal{S} \gets \emptyset$
\State $N_{\text{shared}} \gets 0,\quad N_{\text{dc}} \gets 0,\quad N_{\text{unknown}} \gets 0$

\For{each $q_i \in \mathcal{Q}$}
    \State Determine class of $q_i$ from 5QI
    \If{$q_i \in \{\text{GBR}, \text{Non-GBR}\}$}
        \State $\mathcal{S} \gets \mathcal{S} \cup \{q_i\}$
        \State $N_{\text{shared}} \gets N_{\text{shared}} + 1$
    \ElsIf{$q_i \in \{\text{Delay-Critical GBR}\}$}
        \State $N_{\text{dc}} \gets N_{\text{dc}} + 1$
    \Else
        \State $N_{\text{unknown}} \gets N_{\text{unknown}} + 1$
    \EndIf
\EndFor

\State Compute:
\[
N_{\text{DRB}} \gets
\begin{cases}
1 + N_{\text{dc}} + N_{\text{unknown}}, & N_{\text{shared}} > 0 \\
N_{\text{dc}} + N_{\text{unknown}}, & \text{otherwise}
\end{cases}
\]

\If{$N_{\text{shared}} > 0$}
    \State Create shared DRB $d_s$
    \For{each $q_i \in \mathcal{S}$}
        \State Map $q_i \rightarrow d_s$
    \EndFor
    \State $\mathcal{D} \gets \mathcal{D} \cup \{d_s\}$
\EndIf

\For{each $q_i \in \mathcal{Q}$}
    \If{$q_i \in \{\text{Delay-Critical GBR}, \text{Unknown}\}$}
        \State Create DRB $d_i$
        \State Map $q_i \rightarrow d_i$
        \State $\mathcal{D} \gets \mathcal{D} \cup \{d_i\}$
    \EndIf
\EndFor

\State \Return $\mathcal{D}$

\end{algorithmic}
\end{algorithm}
\vspace{-0.1cm}
\section{Experimental Setup}

\begin{table}[t]
\caption{Radio Access Network Configuration}
\label{tab:ran_config}
\centering
\begin{tabular}{ll}
\toprule
\textbf{Parameter} & \textbf{Value} \\
\midrule
Operating Mode & \textbf{TDD} \\
NR Band & \textbf{n78} \\
Channel Bandwidth & \textbf{40 MHz} \\
Subcarrier Spacing (SCS) & \textbf{30 kHz} \\
Numerology ($\mu$) & \textbf{1} \\
Physical Resource Blocks (PRBs) & \textbf{106} \\
Modulation & \textbf{256-QAM} \\
OAI RF Front-End & \textbf{RF Simulator} \\
RAN Git Commit & \textbf{045140845 } \\
CN Git Commit & \textbf{1f0fc30} \\
TDD Pattern & \textbf{DDDDDDDSUU} \\
Scheduler & \textbf{Proportional-Fair} \\
\bottomrule
\end{tabular}
\end{table}



This section describes the experimental setup used to validate the proposed hybrid QoS-aware DRB mapping framework presented in Section~\ref{sec:sys-arch-prop-approach}. As illustrated in Fig.~\ref{fig:sys-arch}, the experimental testbed consists of three major components: the 5G Core Network~(CN), the Radio Access Network~(gNB), and the User Equipment~(UE). All components are implemented using the OAI platform~\cite{oai_openairinterface5g}. The RAN~(gNB and UE) is built from the \textbf{develop} branch (\textbf{Git commit: {7045140845}} ), while the 5G CN is based on the \textbf{master} branch (\textbf{Git commit: {1f0fc30}}). 

The radio interface is emulated using the \textbf{built-in OAI RF simulator}, which replaces the physical RF front-end while preserving the complete NR protocol stack, including the PHY, MAC, RLC, PDCP, SDAP, RRC, NGAP, and GTP-U layers. This enables repeatable experiments under controlled conditions without requiring software-defined radio hardware. Nevertheless, the proposed implementation can be readily deployed on SDR platforms such as the NI B210 or USRP N310 without modification.

The 5G CN consists of the Unified Data Repository (UDR), Unified Data Management (UDM), Authentication Server Function (AUSF), Network Repository Function (NRF), Access and Mobility Management Function (AMF), Session Management Function (SMF), Policy Control Function (PCF), and User Plane Function (UPF), each deployed as an independent Docker container using the OAI 5G CN framework. Similarly, the modified gNB is deployed using Docker Compose within the same containerized environment. The container image containing the modified gNB implementation is publicly available on DockerHub\footnote{\scriptsize\url{https://hub.docker.com/layers/siriiitdh/miko/gnb-both-qos-types-latest/images/sha256-03e82706e95ef8612f140eed987afa4d0122fd5a24d4a5fbd4d5f6d42d208b2c}}.

The UPF employs the \textbf{OAI eBPF datapath} for user-plane packet forwarding. Packet processing is offloaded to \textbf{in-kernel eBPF programs} attached to the UPF forwarding path, where \textbf{PDRs}, \textbf{FARs}, and \textbf{QERs} are evaluated before forwarding GTP-U packets toward the N3 interface. Consequently, GTP-U decapsulation, QoS flow identification, forwarding decisions, and traffic steering are performed within the \textbf{kernel}, thereby reducing user--kernel context switches while maintaining standards-compliant user-plane processing. The correctness of the eBPF datapath is validated through successful establishment of GTP-U tunnels, correct QoS flow classification, successful end-to-end packet delivery for all configured QoS flows, and verification of the expected QoS-to-DRB mapping during repeated experimental runs.

The gNB and UPF are deployed within the \textbf{same Docker Layer-2 network} to guarantee direct N3 connectivity. Since the eBPF datapath relies on Address Resolution Protocol~(ARP) to resolve the next-hop MAC address before installing forwarding rules, \textbf{Layer-2 adjacency} between the gNB and UPF is required. 
In contrast, deploying the gNB outside the Docker Layer-2 domain prevents successful ARP resolution, thereby inhibiting installation of forwarding rules in the eBPF datapath.

\textbf{To support multiple QoS flows within a single PDU session}, functionality that is not supported by the current OAI implementation at the time of paper writing, several modifications were incorporated across both the control-plane and user-plane protocol stack. These modifications collectively extend the default single-DRB-per-PDU-session implementation to support \textbf{multiple DRBs~(up to 32 DRBs per PDU session)}, thereby enabling hybrid one-to-one and many-to-one QoS flow-to-DRB mapping.

Specifically, the \textbf{NGAP} procedures were extended to transport \textbf{multiple QoS Flow descriptors} during PDU Session Resource Setup and Modification procedures, enabling signalling of multiple QFIs within a single PDU session. The \textbf{RRC} layer was modified to configure \textbf{multiple DRBs} for each PDU session and generate the corresponding \textbf{SDAP} and \textbf{PDCP} configurations required for hybrid bearer establishment. The SDAP layer was enhanced with \textbf{dynamic QFI-to-DRB mapping}, allowing both shared and dedicated DRB assignment according to the proposed hybrid mapping algorithm, together with reverse DRB-to-QFI lookup required for packet forwarding. The PDCP layer was extended to support \textbf{multiple DRB instances} associated with a single PDU session while preserving SDAP-aware packet processing during transmission and reception. The \textbf{E1AP} interface was modified to exchange \textbf{multiple DRB configurations} together with their associated QoS flow information between the CU-CP and CU-UP, ensuring consistency between control-plane signalling and user-plane bearer creation. Finally, the \textbf{GTP-U} processing path was extended to \textbf{preserve QFIs} throughout packet forwarding, enabling end-to-end QoS-aware processing across the SDAP, PDCP, and GTP-U layers.

The gNB and radio configuration used throughout the experiments is summarized in Table~\ref{tab:ran_config}. 



\section{Results and Discussion}

This section validates the implementation of the proposed hybrid QoS-aware DRB mapping framework and evaluates its performance under different traffic conditions. First, the correctness of the implemented QoS Flow-to-DRB mapping mechanism is verified through runtime mapping logs and throughput measurements. Subsequently, the proposed implementation is compared with the baseline OAI implementation under increasing network load.

To validate the implemented mapping mechanism, four QoS Flows were established between the 5G CN and the UE: one GBR flow (5QI~1, QFI~1), one Non-GBR flow (5QI~9, QFI~4), and two Delay-Critical GBR flows (5QI~82, QFI~2 and 5QI~83, QFI~3). The QoS profiles were configured through the OAI QoS framework, while the corresponding QoS parameters were assigned by the SMF.

\begin{figure}[!t]
    \centering
    \includegraphics[width=0.88\columnwidth]{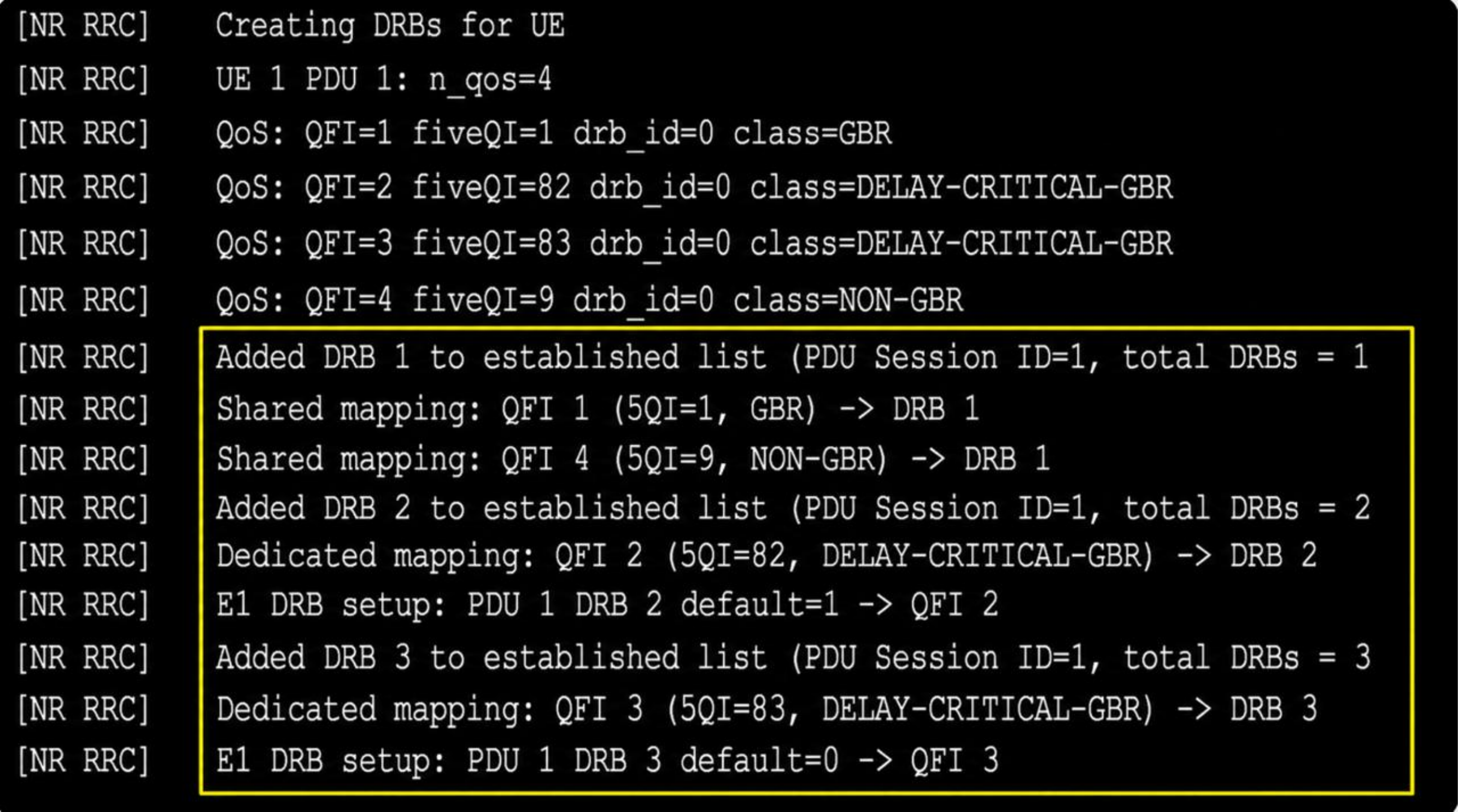}
    \vspace{-0.15cm}
    \caption{Runtime QoS Flow-to-DRB mapping logs at the gNB.}
    \label{fig:qos_logs}
    \vspace{-0.20cm}
\end{figure}

Fig.~\ref{fig:qos_logs} presents the runtime mapping observed at the gNB. The GBR and Non-GBR flows are multiplexed onto a shared DRB (DRB~1), whereas the Delay-Critical GBR flows are assigned dedicated DRBs (DRB~2 and DRB~3). The successful creation of multiple DRBs confirms the correctness of the implemented SDAP-based QoS Flow identification and hybrid mapping mechanism.

To further verify the implementation, simultaneous downlink \textit{iperf} streams were transmitted from the CN to the UE over all configured QoS Flows. Table~\ref{tbl:5QI_Throughput} summarizes the configured QoS parameters and measured throughput. The GBR and Delay-Critical GBR flows operate close to their configured MBR values while satisfying their respective GBR requirements, confirming correct QoS enforcement and service differentiation.

\begin{table}[!t]
\centering
\caption{QoS Throughput Comparison with GBR, MaxBR, and Observed Values}
\label{tbl:5QI_Throughput}
\scriptsize
\setlength{\tabcolsep}{3.2pt}
\renewcommand{\arraystretch}{0.92}
\begin{adjustbox}{max width=\columnwidth}
\begin{tabular}{|c|c|c|c|c|}
\hline
\multicolumn{2}{|c|}{\textbf{5QI Details}} &
\multicolumn{3}{c|}{\textbf{Throughput (Mbps)}}\\
\hhline{~~---}
\textbf{5QI} & \textbf{QoS Type} &
\cellcolor{green!20}\textbf{GBR} &
\cellcolor{blue!15}\textbf{MaxBR} &
\cellcolor{orange!20}\textbf{Actual}\\
\hline
9  & Non-GBR & \cellcolor{green!20}N/A & \cellcolor{blue!15}N/A & \cellcolor{orange!20}N/A\\
\hline
1  & GBR & \cellcolor{green!20}0.512 & \cellcolor{blue!15}3 & \cellcolor{orange!20}2.78\\
\hline
82 & Delay-Critical & \cellcolor{green!20}10 & \cellcolor{blue!15}20 & \cellcolor{orange!20}18.6\\
\hline
83 & Delay-Critical & \cellcolor{green!20}5 & \cellcolor{blue!15}10 & \cellcolor{orange!20}9.28\\
\hline
\end{tabular}
\end{adjustbox}
\vspace{-0.20cm}
\end{table}

Having verified the implementation, the proposed framework is compared with the baseline OAI implementation. Table~\ref{tab:comparison} summarizes the main architectural and performance differences. Multiple QoS-aware DRBs reduce contention among heterogeneous flows, leading to higher aggregate throughput and more efficient use of the fixed radio bandwidth; therefore, the proposed implementation also provides higher aggregate spectral efficiency than the single-DRB baseline.

\begin{table}[!t]
\caption{Performance Comparison Between the Baseline and Proposed OAI Implementations}
\label{tab:comparison}
\centering
\scriptsize
\setlength{\tabcolsep}{3.5pt}
\renewcommand{\arraystretch}{0.90}
\resizebox{\columnwidth}{!}{%
\begin{tabular}{lcc}
\toprule
\textbf{Metric} & \textbf{Baseline} & \textbf{Proposed} \\
\midrule
Multiple DRB Creation & \ding{55} & \ding{51} \\
Aggregate Throughput & Low & High \\
Average Latency & High & Low \\
Aggregate Spectral Efficiency & Low & High \\
QoS Satisfaction & \ding{55} & \ding{51} \\
Runtime QoS Provisioning & \ding{55} & \ding{51} \\
Hybrid QoS Flow-to-DRB Mapping & \ding{55} & \ding{51} \\
\bottomrule
\end{tabular}%
}
\vspace{-0.15cm}
\end{table}

\subsection{Performance Evaluation Methodology}

The proposed framework was evaluated against the baseline OAI implementation using three downlink traffic scenarios. A Delay-Critical GBR flow (5QI~82) was configured with a GBR of 4~Mbps and an MBR of 8~Mbps throughout all experiments. In Scenario~1, the 5QI~82 flow was transmitted together with one 2~Mbps background flow. Scenario~2 added a second 4~Mbps background flow, while Scenario~3 introduced a third 6~Mbps background flow. These scenarios progressively increased contention for radio resources and enabled evaluation of the proposed hybrid DRB mapping under increasing network load. In the baseline implementation, all QoS Flows within a PDU session share a single DRB, whereas the proposed framework maps QoS Flows to shared or dedicated DRBs according to their standardized QoS profiles. This setup evaluates the ability of the proposed framework to preserve the throughput and latency performance of the Delay-Critical GBR flow under increasing traffic load. Each experiment lasted 60~s and was repeated five times. Reported throughput values represent sample means with 95\% confidence intervals, while end-to-end latency was measured for packet sizes ranging from 64 to 1024~bytes.

\subsection{Throughput and Latency Performance Evaluation}

\begin{figure}[!t]
    \centering
    \includegraphics[width=0.94\columnwidth]{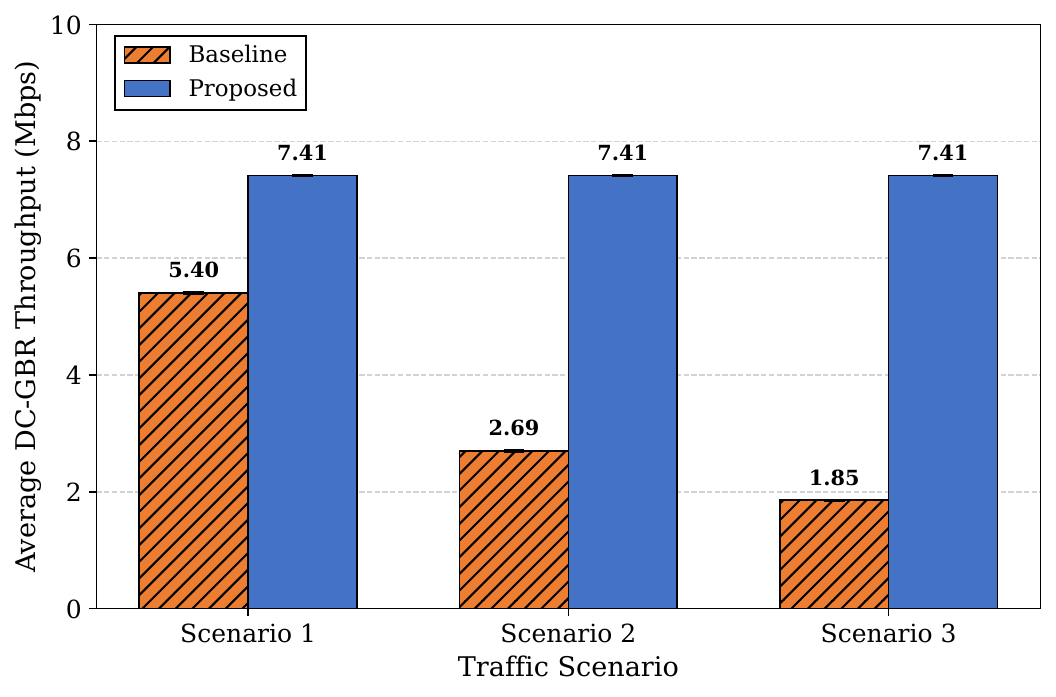}
    \vspace{-0.15cm}
    \caption{Average DC-GBR throughput under the three traffic scenarios}
    \label{fig:throughput_comparison}
    \vspace{-0.20cm}
\end{figure}

Fig.~\ref{fig:throughput_comparison} compares the average throughput achieved by the Delay-Critical GBR flow (5QI~82) under the three traffic scenarios. The proposed implementation consistently maintains a throughput of approximately \textbf{7.41--7.42}~Mbps across all scenarios, whereas the baseline throughput decreases from \textbf{5.40}~Mbps in Scenario~1 to \textbf{2.69}~Mbps and \textbf{1.85}~Mbps in Scenarios~2 and~3, respectively, as additional background traffic increases contention. This behavior results from the proposed hybrid QoS-aware DRB mapping, which assigns QoS Flows to shared or dedicated DRBs according to their standardized 5QI characteristics, thereby protecting the Delay-Critical GBR flow from competing traffic. In contrast, the baseline implementation maps all QoS Flows within a PDU session to a single DRB, causing throughput degradation under increasing load. 

\begin{table}[!t]
\centering
\caption{End-to-End Latency for Different Packet Sizes}
\label{tbl:latency_comparison}

\scriptsize
\setlength{\arrayrulewidth}{0.9pt}
\setlength{\tabcolsep}{2pt}
\renewcommand{\arraystretch}{1.15}

\resizebox{\columnwidth}{!}{%
\begin{tabular}{|
>{\centering\arraybackslash}m{1.85cm}|
>{\centering\arraybackslash}m{2.05cm}|
>{\centering\arraybackslash}m{0.90cm}|
>{\centering\arraybackslash}m{0.90cm}|
>{\centering\arraybackslash}m{0.90cm}|
>{\centering\arraybackslash}m{0.90cm}|
>{\centering\arraybackslash}m{1.00cm}|}
\hline

\multicolumn{2}{|c|}{\textbf{Configuration}} &
\multicolumn{5}{c|}{\textbf{End-to-End Latency (ms)}} \\
\cline{3-7}

\textbf{Scenario} &
\textbf{Implementation} &
\cellcolor{green!15}\textbf{64 B} &
\cellcolor{blue!15}\textbf{128 B} &
\cellcolor{orange!20}\textbf{256 B} &
\cellcolor{purple!15}\textbf{512 B} &
\cellcolor{cyan!20}\textbf{1024 B} \\
\hline

\multirow[c]{2}{1.85cm}{\centering\textbf{Scenario 1}} &
\cellcolor{red!12}\textbf{Baseline} &
\cellcolor{green!8}12.132 &
\cellcolor{blue!8}12.205 &
\cellcolor{orange!10}13.212 &
\cellcolor{purple!8}13.450 &
\cellcolor{cyan!10}14.532 \\
\cline{2-7}

&
\cellcolor{green!18}\textbf{Proposed} &
\cellcolor{green!8}11.628 &
\cellcolor{blue!8}11.789 &
\cellcolor{orange!10}12.150 &
\cellcolor{purple!8}12.560 &
\cellcolor{cyan!10}13.844 \\
\hline

\multirow[c]{2}{1.85cm}{\centering\textbf{Scenario 2}} &
\cellcolor{red!12}\textbf{Baseline} &
\cellcolor{green!8}17.912 &
\cellcolor{blue!8}17.935 &
\cellcolor{orange!10}18.235 &
\cellcolor{purple!8}18.290 &
\cellcolor{cyan!10}18.774 \\
\cline{2-7}

&
\cellcolor{green!18}\textbf{Proposed} &
\cellcolor{green!8}9.946 &
\cellcolor{blue!8}10.408 &
\cellcolor{orange!10}11.195 &
\cellcolor{purple!8}11.327 &
\cellcolor{cyan!10}12.290 \\
\hline

\multirow[c]{2}{1.85cm}{\centering\textbf{Scenario 3}} &
\cellcolor{red!12}\textbf{Baseline} &
\cellcolor{green!8}25.434 &
\cellcolor{blue!8}27.339 &
\cellcolor{orange!10}27.485 &
\cellcolor{purple!8}27.729 &
\cellcolor{cyan!10}27.808 \\
\cline{2-7}

&
\cellcolor{green!18}\textbf{Proposed} &
\cellcolor{green!8}7.870 &
\cellcolor{blue!8}8.919 &
\cellcolor{orange!10}9.896 &
\cellcolor{purple!8}10.188 &
\cellcolor{cyan!10}10.695 \\
\hline

\end{tabular}%
}

\vspace{-0.12cm}
\setlength{\arrayrulewidth}{0.4pt}
\setlength{\tabcolsep}{6pt}
\renewcommand{\arraystretch}{1}

\end{table}

Table~\ref{tbl:latency_comparison} further shows that, although latency increases slightly with packet size for both implementations, the baseline experiences a significant increase in latency as the number of background flows increases. The proposed framework, however, maintains consistently low latency across all scenarios. Overall, the throughput and latency results demonstrate effective traffic isolation and QoS preservation without requiring modifications to the underlying radio scheduler.

\subsection{Scalability Evaluation}

To evaluate scalability, one GBR flow (5QI~1) and one Delay-Critical GBR flow (5QI~82) were transmitted simultaneously. The offered GBR load was increased from 40 to 140~Mbps, while the Delay-Critical flow retained its original QoS configuration. Each configuration ran for 360~s and was repeated ten times.

\begin{figure}[!t]
    \centering
    \includegraphics[width=0.88\columnwidth]{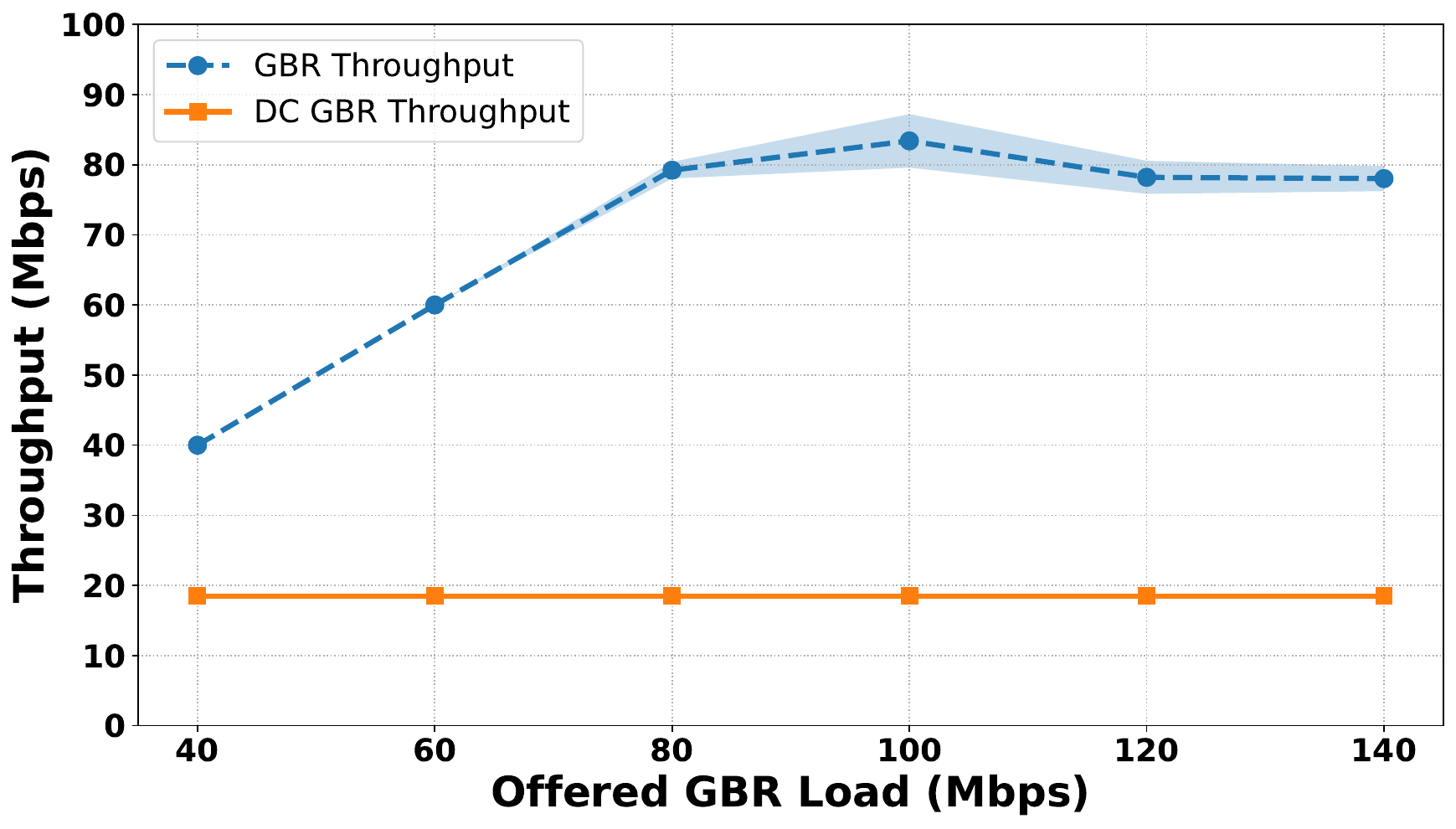}
    \vspace{-0.15cm}
    \caption{Impact of increasing GBR load on Delay-Critical GBR throughput.}
    \label{fig:gbr_throughput_plot}
    \vspace{-0.20cm}
\end{figure}

As shown in Fig.~\ref{fig:gbr_throughput_plot}, GBR throughput initially increases with offered load and then saturates as the available radio resources are fully utilized. In contrast, the Delay-Critical GBR throughput remains nearly constant and close to its configured MBR. This confirms that the proposed framework preserves higher-priority traffic despite increasing load from the competing GBR service.

\subsection{Discussion}

The runtime logs and throughput measurements validate the correct realization of the hybrid QoS Flow-to-DRB mapping mechanism. Compared with the baseline single-DRB implementation, dynamically creating shared and dedicated DRBs reduces contention, preserves GBR throughput, lowers latency, and maintains Delay-Critical GBR performance near the capacity limit of the testbed. The proposed OAI implementation therefore provides practical QoS differentiation for heterogeneous services within a single PDU session while remaining compliant with the standardized 3GPP QoS framework.

\vspace{-0.6em}
\section*{Acknowledgments}
This work is partly supported by NSF grant CNS-1956357, and partially supported by the project Synergy:
Taking openness to the next level in 6G networks. 

 \section{Summary and Future Works}
 
 The proposed work presents a hybrid QoS-aware DRB mapping mechanism that dynamically assigns QoS flows to shared or dedicated DRBs based on standardized 5QI characteristics. Implemented on the OpenAirInterface platform, it supports multiple QoS flows within a single PDU session while improving radio resource utilization. Experimental results validate the mapping and demonstrate improved throughput for GBR traffic together with traffic isolation for Delay-Critical flows. Overall, the proposed framework provides a practical and scalable solution for QoS-aware radio resource allocation in B5G networks. Future work will focus on dynamic QoS flow management through PFCP Session Modification for runtime addition and removal of QoS flows.

\bibliographystyle{IEEEtran}
\bibliography{ref}

@techreport{3gpp_ts23501_v16_6_0,
  title        = {{System Architecture for the 5G System (5GS)}},
  institution  = {3rd Generation Partnership Project (3GPP)},
  number       = {TS 23.501},
  version      = {16.6.0},
  year         = {2020},
  note         = {Release 16}
}

@misc{oai_openairinterface5g,
  title        = {{OpenAirInterface 5G: Open-Source Software for 5G Wireless Systems}},
  author       = {{OpenAirInterface Software Alliance}},
  year         = {2015},
  howpublished = {\url{https://gitlab.eurecom.fr/oai/openairinterface5g}},
  note         = {Accessed: 2026-04-08}
}

@inproceedings{roughan2004class,
  title={{Class-of-Service Mapping for QoS: A Statistical Signature-based Approach to IP Traffic Classification}},
  author={Roughan, Matthew and Sen, Subhabrata and Spatscheck, Oliver and Duffield, Nick},
  booktitle={Proceedings of the 4th ACM SIGCOMM conference on Internet measurement},
  pages={135--148},
  year={2004}
}

@inproceedings{satka2022qos,
  title={{QoS-MAN: A Novel QoS Mapping Algorithm for TSN-5G Flows}},
  author={Satka, Zenepe and Ashjaei, Mohammad and Fotouhi, Hossein and Daneshtalab, Masoud and Sj{\"o}din, Mikael and Mubeen, Saad},
  booktitle={2022 IEEE 28th International Conference on Embedded and Real-Time Computing Systems and Applications (RTCSA)},
  pages={220--227},
  year={2022},
  organization={IEEE}
}

@inproceedings{zhang2021towards,
  title={{Towards a Sotware-Defined, Fine-Grained QoS Framework for 5G and Beyond Networks}},
  author={Zhang, Zhi-Li and Dayalan, Udhaya Kumar and Ramadan, Eman and Salo, Timothy J},
  booktitle={Proceedings of the ACM SIGCOMM 2021 Workshop on Network-Application Integration},
  pages={7--13},
  year={2021}
}

\end{document}